\documentclass[aps,pre,amsmath,amssymb,amsfonts,twocolumn,dvipdfmx]{revtex4-1}

\usepackage{graphicx}
\usepackage{dcolumn}
\usepackage{bm}
\usepackage{epsfig,grffile,tipa}

\begin{document}

\preprint{APS/123-QED}

\title{Monte Carlo study of the frame, fluctuation and internal tensions of fluctuating membranes with fixed area}

\author{Hayato Shiba}
\affiliation{Institute for Solid State Physics, University of Tokyo, Chiba 277-8581, Japan}
\author{Hiroshi Noguchi}
\affiliation{Institute for Solid State Physics, University of Tokyo, Chiba 277-8581, Japan}
\author{Jean-Baptiste Fournier}
\affiliation{Laboratoire Mati\`ere et Syst\`emes  Complexes (MSC), UMR 7057 CNRS, Universit\'e Paris Diderot, F-75205, Paris, France}

\date{\today}

\begin{abstract}
Three types of surface tensions can be defined for lipid membranes: the internal tension, $\sigma$, conjugated to the real membrane area in the Hamiltonian, the mechanical frame tension, $\tau$, conjugated to the projected area,  and the ``fluctuation tension'', $r$, obtained from the fluctuation spectrum of the membrane height. 
 We investigate these surface tensions by means of a Monge gauge lattice Monte Carlo simulation involving the exact, nonlinear, Helfrich Hamiltonian and a measure correction for excess entropy of the Monge gauge. Our results for the relation between $\sigma$ and $\tau$ agrees well with 
the theoretical prediction of [J.-B. Fournier and C. Barbetta, Phys. Rev. Lett., 2008, 100, 078103] based on a Gaussian approximation. This provides a valuable knowledge of~$\tau$ in the standard Gaussian models where the tension is controlled by $\sigma$.
However, contrary to the conjecture in the above paper, we find that $r$ exhibits no significant difference from $\tau$ over more than five decades of tension. Our results appear to be valid in the thermodynamic limit and are robust to changing the ensemble in which the membrane area is controlled.
\end{abstract}

\pacs{Valid PACS appear here}
\maketitle


\section{Introduction}
A few decades of research has clarified that lipid bilayers can be treated as sheets obeying the continuum curvature/tension elasticity formulated by Helfrich\cite{Helfrich1973}. Characterising membrane elastic constants in numerical simulations has become an important issue  due to the increase in computational power and  development of simulation methods.  Starting with the seminal work of Ref.\citenum{Goetz1999}, 
a growing number of research papers are dedicated
to the determination of bending rigidity\cite{Lindahl2000,denOtter2003,Farago2004,Harmandaris2006,Noguchi2006,
Shiba2011,Noguchi2011,Hu2013,Tarazona2013,Watson2013}, Gaussian modulus\cite{Hu2012}, and surface tension\cite{Imparato2006,Noguchi2006,Shiba2011,Farago2011,Tarazona2013}, from   molecular simulation of all-atom systems 
to coarse-grained models. 

A number of questions have been raised 
regarding the definition of membrane surface tension. 
When a thermally fluctuating flexible membrane is stretched
by a moderate lateral tension, the membrane
extends its projected area by suppressing its transverse
fluctuation\cite{Fournier2001}.
This leads to multiple definitions of membrane tension:
i) ``internal tension'', $\sigma$, conjugated
to the real area $A$ in the membrane Hamiltonian, ii) mechanical ``frame tension'', $\tau$,
conjugated to the projected area $A_{\rm p}$,
\cite{David1991} and iii) ``fluctuation tension'', $r$, associated to the lowest-order wavevector dependence of the inverse fluctuation spectrum. The latter two are directly experimentally measurable\cite{Sengupta2010}.  

Some discussions on the difference between these
surface tensions have been lasting for more than
a decade. First of all, it is  well accepted that $\sigma$ and $\tau$ are intrinsically different, because $\sigma$ is a microscopic parameter belonging to the effective Hamiltonian while $\tau$ is a macroscopic observable integrating the local stress tensor over all the membrane fluctuations. Based on free-energy calculations and stress tensor averages, the difference between $\sigma$ and $\tau$ has been estimated within the Gaussian approximation~\cite{Farago2003,Fournier2008,Imparato2006}. 
Note that the validity of this estimation was questioned by Schmid~\cite{Schmid2011a}. 
It is also well accepted that $\sigma$ and $r$ are different, because $r$ is actually the renormalised version of $\sigma$ (at the Gaussian level they are equal). The relationship between $r$ and $\tau$ has been much debated. Several authors have argued theoretically\cite{Cai1994a,Farago2004,Farago2011,Diamant2011}, or observed in numerical simulations\cite{Goetz1999,Lindahl2000,denOtter2003,Farago2011,Noguchi2006,Shiba2011,Marrink2001,Wang2005,Brannigan2006,West2009,Neder2010,Schmid2011a}, that $r$ matches the mechanical frame tension $\tau$. Other authors have observed a difference between $r$ and $\tau$, either in numerical simulations
\cite{Imparato2006,Fournier2008,Tarazona2013,Stecki2006} or experiments\cite{Sengupta2010}. 

These questions should be carefully revisited with the help of 
extensive, large-scale numerical simulations. A complication has to do with the equivalence of 
ensembles in the thermodynamic sense, as in the simulations found in the literature, either the 
areas $A$ and $A_{\rm p}$ or their conjugated tensions $\sigma$ and $\tau$ are fixed. Different ensembles are only equivalent in the thermodynamic limit of 
infinite membranes, $A\to\infty$ at fixed $A/N$, with $N$ the total number of degrees of freedom.
It is therefore very important to check the convergence of the numerical results with system size.
In another recent simulation\cite{Avital2015}, $r$ was claimed to deviate from~$\tau$ for $\tau<0$ 
and to coincide with~$\tau$ for $\tau>0$. 
It is thus interesting to study how $r$ behaves when $\tau$ assumes positive values very close to zero, for which the system undergoes large fluctuations.
Let us stress that investigating several orders of magnitude of $\tau$ is important, because 
membrane tension can experimentally vary over more than five decades\cite{Evans1990}. 

In this paper, we  investigate the differences between the three surface 
tensions, $\sigma$, $\tau$ and $r$, over several decades of $\tau$, by  
means of Monte Carlo simulations of a lattice membrane system in the Monge gauge. 
In the lattice model, the internal tension $\sigma$ and the frame tension $
\tau$ can be controlled and
short range fluctuations generated by molecular or particle protrusions, which make accurate 
estimations of $r$ difficult, are not present. 
However, a measure correction is required to cure the excess entropy due to the unidirectional lattice-site motions implied by the Monge gauge.
We also employ exact nonlinear expressions to calculate the membrane's curvature and area. 
We find that $\sigma$ differs from $\tau$ at small positive frame tensions,
while they match at moderate and high frame tensions. Our results for $\sigma-\tau$ agree well with the prediction of ref.~\citenum{Fournier2008} based on a Gaussian approximation, although our simulation investigates regimes where the Gaussian approximation is far from being justified.
We find also that $r$ and $\tau$ match, within the accuracy of our simulation, over more than five decades, even in the limit of very small frame tensions.
These relationships remain in the  thermodynamic limit and seem not to depend on the ensemble in which the real area $A$ is controlled
 (see Appendix~\ref{ensembleAfixed}).

Our paper is organised as follows. In Sec.~\ref{sec:model}, we detail our lattice membrane model and the Monte Carlo simulation framework that realises an equilibrium ensemble for fixed $\tau$ and $\sigma$. In this  framework, the average of the real membrane area, $\langle A\rangle$, is kept constant by adjusting $\sigma$ while $\tau$ is varied. We also perform our simulation in the equilibrium ensemble for fixed $\tau$ and $A$. This is discussed in Appendix~\ref{ensembleAfixed}. In the thermodynamic limit of large systems, we thus simulate a membrane with fixed real area subjected to a variable frame tension. In Sec.~\ref{sec:results}, we numerically determine and we compare the three tensions $\sigma$, $\tau$ and~$r$. We give our conclusions in Sec.~\ref{sec:concl}.

\section{Model and method\label{sec:model}}
\subsection{Stress-controlled ensemble}

Let us consider an incompressible, fluctuating lipid membrane with a fixed area $A=A_0$, corresponding to a fixed number of lipids. We assume that it is attached to a deformable frame, of variable area $A_{\rm p}$, that exerts onto the membrane a fixed tension~$\tau$. The Hamiltonian of the system is thus given by\cite{Helfrich1973} 
\begin{equation}
\mathcal{H} = -\tau A_{\rm p} +  \int_{A}\,\frac{\kappa}{2}H^2\,dA,\quad(\tau, A\mathrm{~fixed;~}A_{\rm p}\mathrm{~free})
\label{eq:sah}
\end{equation}
where $H=c_1+c_2$ represents the sum of the two principal curvatures of the membrane. The Gaussian bending modulus\cite{Helfrich1973} $\bar\kappa$ has been neglected, according to the Gauss--Bonnet theorem, which is correct for fixed angular boundary conditions or periodic boundary conditions along the frame\cite{Buchins_book}.

We use the Monge representation, in which the membrane shape is describe by its height $z = h(x,y)$ above the plane of the frame. 
Note that this parametrisation does not require the membrane deformation to be small. The only restriction is that overhangs are forbidden, since the function $h$ is single valued. In this representation, the total curvature $H$ is exactly given by\cite{Safran_book}
\begin{equation}
H = \frac{(1+h_x^2) h_{yy} + (1+h_y^2) h_{xx} - 2h_x h_y h_{xy} }{ ( 1+h_x^2+h_y^2 )^{3/2}}, \label{eq:prec}
\end{equation}
where subscripts represent spatial derivatives, e.g., $h_x = \partial h/\partial x$.
We work here in the non-linear regime, allowing for large deformations of the membrane, i.e., we do not use the custom linear approximation $H\simeq h_{xx}+h_{yy}$ (employed in the Gaussian framework). Note that in our simulations (see the details below)  the maximum slope, $\mathrm{max}(|h_x|)$, reaches 1.3 at vanishing frame tension, but does not exceed this value. This justifies both using the exact expression for $H$ and  neglecting overhangs. Close to the buckling transition, however, i.e., for large negative frame tensions, these assumptions may not be valid.

Because it is quite difficult to work with a fixed area $A=A_0$, we change ensemble in order to control the conjugated internal tension $\sigma$, instead of the membrane area $A$. In the thermodynamic limit, the two ensembles are expected to be equivalent.  Indeed, as shown in Appendix~\ref{ensembleAfixed}, we find quantitatively similar results in the ensemble in which the real area $A$ is almost prescribed by means of a quadratic potential.
In this stress--controlled ensemble, the modified Hamiltonian weighting the membrane fluctuations is given by
\begin{equation}
\mathcal{H}^\star =  - \tau A_{\rm p} + \int_{A}\,\frac{\kappa}{2}H^2\,dA+\sigma A,\quad(\tau,\sigma\mathrm{~fixed;~}A_{\rm p}, A\mathrm{~free})
\label{eq:Hamil1}
\end{equation}
with $\langle A\rangle=A_0$ systematically enforced by adjusting $\sigma$.

\subsection{Lattice model}

In order to implement the membrane fluctuations numerically, 
we introduce a $N_x\times N_y$ lattice on the plane $(x,y)$ of the frame, with lattice spacing $a$. 
The height of the membrane surface is defined  by $h_{ij}$ on each lattice site, with $1\le i\le N_x$ and $1\le j\le N_y$. 
In the reminder of this paper, we take $N_x=N_y=\sqrt{N}$. Furthermore, we adopt periodic boundary condition. Thus the projected area is  
given by $A_{\rm p} = Na^2$.

Importantly, because the number of lipids is fixed while the frame area $A_{\rm p}$ is variable, we take $N$ constant but $a$ variable. In other words, in our simulation the lattice spacing is not fixed.  We stress, however, that $a$ always remains {\em uniform} over the lattice: it is the overall lattice spacing that changes. Therefore the projection over the reference plane of the vertices always maps onto a square grid of finite spacing and the vertices can never overlap. Note that 
a similar lattice membrane model (with fixed spacing  and the Gaussian approximation) has been employed for Monte Carlo simulation 
in order to investigate inclusions effects~\cite{Weikl2001}.

Monte Carlo simulations are performed
with the lattice heights $\{h_{ij}\}$ and the lattice spacing $a$ as dynamical variables. Special care is required, however, for the sampling of the height variables. In the Monge gauge, each site can only fluctuate in the $z$ direction, whereas in the physical gauge, because the membrane is fluid, the relevant fluctuations occur by local displacements along the membrane's normal. In order to avoid an artificial entropy increase of the states where the membrane is tilted, the naive measure $d^N[h]=\prod_{i,j}dh_{ij}/\delta_0$, where $\delta_0$ is a quantum of height fluctuations, must be transformed to
\begin{align}
\prod_{i,j}\frac{dh_{ij}}{\delta_0/\cos\theta_{ij}}.
\end{align}
Here, $\theta_{ij}$ is the angle between the normal vector of the site $(i,j)$ and the $z$ axis, and $\cos \theta_{ij} = 1/\sqrt{1+h_x^2+h_y^2}$.
In this way, the quantum height displacement in the normal direction is always $\delta_0$. Equivalently, we can keep the naive measure $d^N[h]$ and add to the Hamiltonian the following correction term:
\begin{equation}  
\mathcal{H}_{\rm corr} = - k_{\rm B}T \sum_{i,j} \ln (\cos\theta_{ij})\,,
\label{eq:Hcor}
\end{equation}
where $T$ is the temperature and $k_\mathrm{B}$ Boltzmann's constant.
Note that this correction, that would be exact if the membrane was uniformly tilted, is only approximate if the tilt changes from site to site. In practice, it works well when the membrane slope is small, as in the present simulations.
For highly tilted membranes, such as buckled membranes, 
the correction term, eqn~(\ref{eq:Hcor}), is not sufficient, as discussed in Appendix~\ref{app:ang}.

The partition function in the stress--controlled ensemble introduced above is given by
\begin{equation}
Z_N (\tau,\sigma) = \int dA\int dA_{\rm p} \int d^N[h]\ \exp [-\mathcal{H}'/(k_\mathrm{B}T) ],
\end{equation}
with $\mathcal{H}'=\mathcal{H^\star}+\mathcal{H}_{\rm {corr}}$.
Metropolis sampling is employed~\cite{Landau2009}.
In each Monte Carlo step, $N$ trial moves of the height of a random site
are attempted.
The change of $a$ (hence $A_{\rm p}$) is attempted every $5$ Monte Carlo steps. The amplitudes of the $h_{ij}$ and $A_{\rm p}$ changes are adjusted 
so that the rejection rate lies in the range 40\%--60\%. 
Both $A$ and $A_{\rm p}$ change during the simulations. However, for a given Monte Carlo run at a specified value of $\tau$, the internal tension $\sigma$ is adjusted in order for the average of the membrane area $\langle A\rangle$ to have the specific value $A_0$, fixed once for all. 

In our simulation, the real area $A$ of the membrane is calculated as follows. 
The local area associated to a site $P(i,j)$ is obtained as
one half of the total area of  the four triangles [APB], [BPC], [CPD] and [DPA] built from the neighbouring sites of coordinates:
\begin{equation}
{\rm A} (i+1,j),\ {\rm B}(i,j+1),\ {\rm C}(i-1,j),\ {\rm D}(i,j-1).
\label{eq:Ai}
\end{equation}
The area of each triangle can be calculated 
by using height differences between the lattice sites. 
Then, the total real area can be expressed as
\begin{equation}
A = \sum_{i=1}^N \sum_{j=1}^N \frac{1}{2}(T_{\rm {APB}} + T_{\rm {BPC}} + T_{\rm {CPD}} + T_{\rm {DPA}}). 
\label{eq:Asum}
\end{equation}

Simulation are carried out 
at least for $10^7$ 
steps after the initial relaxation.
The statistical errors are calculated from four or more independent runs.

\subsection{Dimensionless units and orders of magnitudes}
\label{sec:dimensionless}

Since our simulated membrane has a prescribed \textit{average} area $A_0$ and a fixed number $N$ of degrees of freedom, we can picture each degree of freedom as a patch of lipids of fixed average area $a_0^2=A_0/N$. In the following we are going to nondimensionalise the lengths by $a_0$ (the size of a degree of freedom in the membrane) and the energies by $k_\mathrm{B}T=4\times10^{-21}$~J (room temperature energy). In other words, we shall set $a_0=k_\mathrm{B}T=1$. Hence, all tensions will be given in units of $k_\mathrm{B}T/a_0^2$. As previously discussed, we shall thus adjust $\sigma$ for each $\tau$ in order to have systematically $\langle A\rangle= N$, in dimensionless units.

\begin{figure}
\centering
 \includegraphics[width=.9\linewidth]{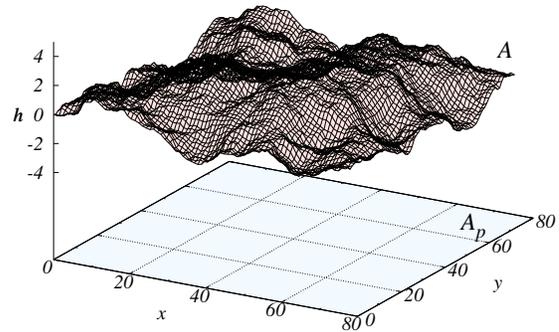}
  \caption{Simulation snapshot for $\kappa=10~k_\mathrm{B}T$, $\tau=0$ (tensionless frame), $N=6400$, and $\langle A\rangle=Na_0^2$ achieved for $\sigma\simeq0.50~k_\mathrm{B}T/a_0^2$. The lengths are given in units of $a_0$, which corresponds to the size of a coarse-grained degree of freedom in the membrane. In this simulation, the equilibrium lattice spacing is $a\simeq0.983\,a_0$, which corresponds to $(A-A_{\rm p})/A\simeq 0.034$. 
With typically $a_0\approx20~\text{nm}$ and $k_\mathrm{B}T=4\times10^{-21}$~J, the residual internal tension is about $\sigma\simeq5\times10^{-6}$~J/m$^2$.}
\label{fig:view3d}
\end{figure}

Let us estimate tension scale $k_\mathrm{B}T/a_0^2$. In principle, we are free to choose the linear size $a_0$ of the lipid patches that constitute our degrees of freedom. It is convenient, however, to choose the smallest possible size, in order to allow for all possible fluctuations and to avoid using renormalised elastic constants. Owing to the level of coarse-graining of our simulation, in which the bilayer membrane of real thickness $\approx\!5$~nm is treated as a mathematical surface, it is natural to think of the lipid patches as regions of typical size, e.g. $a_0\approx20~\text{nm}$. Then, the natural unit of tension in our simulation is $k_{\mathrm B}T /a_0^2\approx10^{-5}$~J/m$^2$. The tension mechanically imposed on ordinary vesicle membranes, in experiments, are in the range $10^{-7}$~J/m$^2$--$10^{-2}$~J/m$^2$, where the smallest value corresponds to floppy vesicles and the largest value corresponds to the lytic tension\cite{Rawicz2000}. We shall therefore have $\tau$ span the range $10^{-2}$--$10^{3}$ in dimensionless units. Let us stress that this corresponds to five orders of magnitude.

In our simulation, we choose a specific value of the bending rigidity corresponding to $\kappa=10~k_\mathrm{B}T$, i.e., $\kappa=10$ in dimensionless units.
We shall consider membrane sizes corresponding to $N=400,1600, 6400$ and $25600$. With the value of $a_0$ given above, this corresponds to membrane of linear dimension $\sqrt{A_0}=\sqrt{N}a_0=400$~nm, $800$~nm, $1.6~\mu$m and $3.2~\mu$m, respectively. These values are quite small if one thinks of macroscopic experiments, but they are reasonably large on the biological scale. Because of the change of ensemble that we have performed, we shall check the convergence of our results with increasing $N$.

In Fig.~\ref{fig:view3d}, a simulation snapshot for $N=6400$ is shown in the tensionless state $\tau=0$.

\section{Results\label{sec:results}}

\subsection{Frame tension and internal tension}

From now, unless otherwise specified, all quantities will be given in dimensionless units (see Sec.~\ref{sec:dimensionless}). We investigate here the relationship between $\sigma$ and $\tau$. 
As explained above, we place ourselves in the $(\sigma,\tau)$ ensemble; however, for every imposed frame tension $\tau$, 
we determine the internal tension $\sigma$ that achieves $\langle  A\rangle = N$. Our simulated membrane thus effectively has a constant real area (up to fluctuations that become irrelevant in the thermodynamic limit). Simulations in the $(A,\tau)$ ensemble, where $A$ is nearly fixed by means of a steep quadratic potential, are discussed in Appendix~\ref{ensembleAfixed}.

\begin{figure}
\centering
 \includegraphics[width=.85\linewidth]{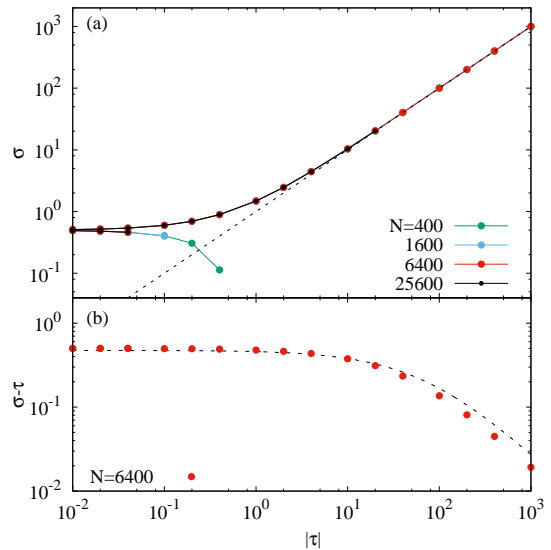}
  \caption{(a) The values of $\sigma$ corresponding to the prescribed average membrane area $\langle A\rangle = N$ are plotted against $|\tau|$, for system sizes $N=400, 1600, 6400$ and $25600$. The upper (resp.\ lower) branch displays the data for $\tau>0$ (resp.\ $\tau<0$). For $\tau<0$ the data points are plotted up to the buckling transition. The dotted line indicates $\sigma =\tau$ (for $\tau >0$). (b) The red points show the difference $\sigma-\tau$ as a function of $\tau>0$ for $N=6400$; the black dashed line shows the best fit to the theoretical Gaussian prediction eqn~(\ref{eq:tsdiff4})--(\ref{eq:cutoffalpha}).}
\label{fig:tsplot}
\end{figure}

The determined values of $\sigma$ are plotted in Fig.~\ref{fig:tsplot}a against the frame tension for different system sizes: $N=400, 1600, 6400$ and $25600$. For any value of $\tau$ we find $\sigma>\tau$. Data for both positive and negative frame tensions 
are displayed together as a function of $|\tau|$: the upper branch corresponds to $\tau>0$ and the lower branch corresponds to $\tau<0$. Note that for $\tau>0$ the difference between $\sigma$ and $\tau$ is sizeable only for small values of the frame tension. For $\tau=0$ we find that the residual internal tension is equal to $\sigma_0\simeq0.500\pm0.001$, where the error bar takes into account the determinations using different values of $N$. In other words, the residual tension turns out to be fairly independent of system size.

\begin{figure}
\centering
 \includegraphics[width=.85\linewidth]{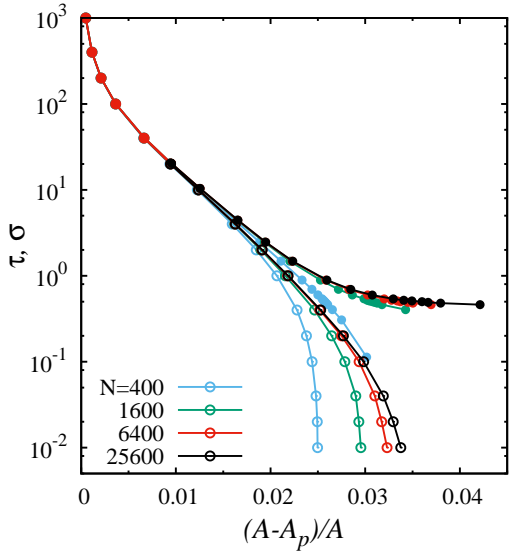}
  \caption{The internal tension $\sigma$ (closed circles) and the frame tension $\tau$ (open circles) are plotted as a function of residual area $(A-A_{\rm p})/A$ for various system sizes $N=400, 1600, 6400$ and $25600$. The data for $\tau$ is not shown in the region $\tau<0$. Finite size effects becomes small as the system size increases: the data suggest that the thermodynamic limit is almost reached for $N=6400$. All the statistical errors are small, within the symbol marks.
  }
\label{fig:tsra2}
\end{figure}

When $\tau$ decreases below some negative threshold $\tau_b<0$, the membrane buckles either into the $x$ or the $y$ direction: the average membrane shape undergoes a transition from flat to non-flat through a symmetry breaking (like a rod under compression). Note that the membrane height function remains single valued although it acquires a bimodal distribution.
The absolute value $|\tau_b|$ of the buckling threshold decreases as the system size $N$ gets larger. In the lower branch, the data points for $|\tau| > |\tau_b|$, in the buckling state, are eliminated from the plots  in Figs.~\ref{fig:tsplot}a and \ref{fig:tsra2}. 
It is notable that the condition $\sigma <0$ 
is not required for the buckling transition. We also simulated our lattice model in the $(A_{\rm p}, \sigma)$ ensemble, i.e. with constant lattice spacing $a$ and freely changing membrane area $A$. We found that the membrane buckles then at $\sigma=0$. 
These results suggest that $\sigma$ is not the mechanical force that drives membrane buckling.

In Ref.\citenum{Fournier2008}, the difference between the tensions $\sigma$ and $\tau$ was calculated within the Gaussian approximation, yielding
\begin{equation}
\label{eq:tsdiff4}
\sigma-\tau = \frac{k_\mathrm{B}T\Lambda^2}{8\pi} 
\left[ 1-\frac{\sigma}{\kappa\Lambda^2} \ln \left( 1+\frac{\kappa\Lambda^2}{\sigma} \right) \right],
\end{equation}
where $\Lambda$ is the ultraviolet wavevector cutoff and all quantities have their normal dimensions. Within our simulation, this ultraviolet cutoff corresponds to the lattice spacing. We thus expect
\begin{equation}
\Lambda=\alpha\,\frac{\pi}{a}\,,
\quad\text{with~}\alpha\approx1\,.
\label{eq:cutoffalpha}
\end{equation}
Whereas our simulation is fully nonlinear, the expression for $\sigma-\tau$ calculated within the Gaussian approximation fits the data quite well (Fig.~\ref{fig:tsplot}b). 
Since the numerical value of $\langle a\rangle$ changes by only 
$1.4$\% from $\tau=0$ to $1000$, we use for the fit 
the extreme value at high tension, $a = 1$. 
The best fit of the data gives then $\alpha = 1.12$, very close to unity.

\textit{Thermodynamic limit.}---In Fig.~\ref{fig:tsra2}, we plot the two tensions $\sigma$ and $\tau$ against the excess area $(A-A_{\rm p})/A$ for the various system sizes $N=400, 1600, 6400$ and $25600$.
Here, only positive values of $\tau$ are displayed: the downwards divergence of the  curves associated to the plain circles corresponds to $\tau\to0$. The system size dependence is marked for small sizes, but the data appears to converge to a universal thermodynamic limit at large system sizes.
Finite size effects are more or less irrelevant for $N\ge 6400$.

\subsection{Fluctuation tension \label{sec:calc_r}}

\begin{figure}[t]
\centering
 \includegraphics[width=.75\linewidth]{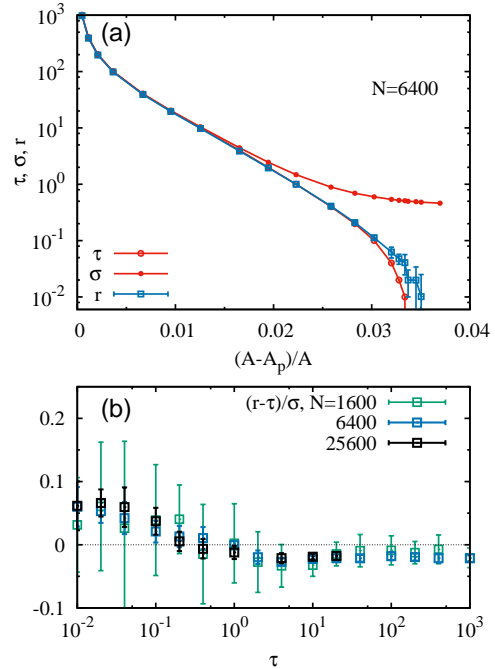}
  \caption{(a) Plot of $\tau,\ r$, and $\sigma$ as a function of excess area $(A-A_{\rm p})/A$ for $N=6400$, which roughly corresponds to the thermodynamic limit. The data for $\tau$ and $\sigma$ are the same as in Fig.~\ref{fig:tsra2},  and $r$ is plotted with the error bars. As for the statistical errors in $(A-A_{\rm p})/A$, they are small, within the symbol marks.   
(b) The difference between $r$ and $\tau$, normalised by $\sigma$, is plotted against $\tau$ for $N= 1600$, $6400$, and $25600$. The convergence of the data with increasing $N$ indicates that the thermodynamic limit is reached for $N=6400$. For the sake of comparison, a plot of $(\sigma -\tau)/\sigma$ is also shown.
}
\label{fig:ren} 
\end{figure}

We now discuss the fluctuation tension $r$. 
It is derived from the 
spectrum density, assumed to take the standard form\cite{Safran_book}:
\begin{equation}\label{eq:spectrum}
\langle | h(\bm q)|^2 \rangle = \frac{k_\mathrm{B}T}{rq^2 + \kappa_r q^4},
\end{equation}
where $h(\bm q)$ is the Fourier transform of the height function, with $\bm q$ the wavevector. The fluctuation tension $r$ and the bending rigidity $\kappa_r$, appearing in the fluctuation spectrum, are the renormalised counterparts of the Hamiltonian parameters $\sigma$ and $\kappa$.

In estimating these quantities using simulation results, it is crucially important to consider large-scale fluctuations. However, because we always treat finite-size systems, the wavevector spectrum is limited, which results in systematic deviations. Previously, two of the present authors have been addressed in detail the question of the estimation 
for the bending rigidity from simulation data\cite{Shiba2011}. For a planar membrane, the correct value of the bending rigidity can be determined uniquely by first estimating the rigidity $\kappa_r$ via a fit of the inverse spectrum $(rq^2+\kappa_rq^4)/(k_\mathrm{B}T)$ in the wavevector range $2\pi/( a_0\sqrt{N}) \le q < q_\mathrm{cut}$,
then by extrapolating the results to the limit $q_{\rm cut}\to0$, where $q_\mathrm{cut}$ is a varying upper fitting wavevector limit.
Note that because the projected area $A_{\rm p}$ fluctuates, there is some ambiguity regarding the definition of the quantified wavevectors. As discussed in Appendix~\ref{wavevectors}, our results are not sensitive to those details. This procedure has been carried out for the present system. We find that the value of $\kappa_r$ coincides with the bare value $\kappa$, up to our numerical precision. This result makes sense because
renormalisation group calculations predict
\begin{equation} \label{eq:dkappa}
\Delta\kappa=\kappa_r-\kappa\approx
-\frac{k_\mathrm{B}T}{8\pi}\,\ln N\,,
\end{equation}
up to a numerical factor of order unity\cite{Safran_book} (here all quantities have their normal dimensions). This prediction gives $\Delta\kappa/\kappa\approx0.03$ for $N=6400$, which is less than the error bars, and thus not detectable in our simulation.

The correct value of the fluctuation tension $r$ (the renormalised tension) can be estimated likewise in the limit $q_{\rm cut}\to 0$.
For finite values of $q_{\rm cut}$, we find that $r$ is overestimated but converges in the limit  $q_{\rm cut}\to0$ to a well defined value. The corresponding extrapolation is performed by using a quadratic function of $q_{\rm cut}$. The fitting was performed in the range $ 0.015 \le (q_{\rm cut}/\pi )^2 < 0.15$. 

Figure~\ref{fig:ren} shows the values of the fluctuation tension $r$. 
We find that $r$ is fairly close to $\tau$ for all the investigated values of $\tau$, as indicated in Fig. \ref{fig:ren}a. 
In Fig.~\ref{fig:ren}b, the difference $r-\tau$ is plotted against $\tau$. For large system sizes, $N=6400$ and $25600$, we observe a small deviation between $r$ and $\tau$; however, as discussed in Appendix~\ref{app:ang}, we believe that this deviation is smaller than a possible systematic error arising from the measure correction.

\subsection{Comparison with the Gaussian approximation}
\begin{figure}[t]
\centering
 \includegraphics[width=.85\linewidth]{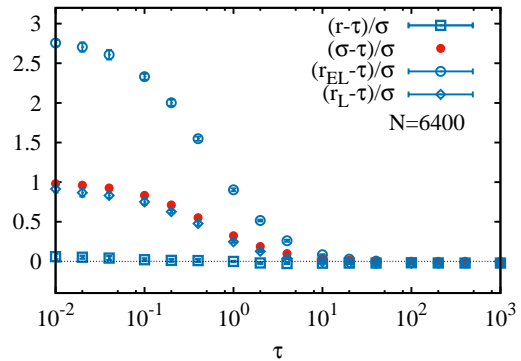}
\caption{
Excess fluctuation tensions $r$, $r_{\rm EL}$ and $r_{\rm L}$, measured with respect to $\tau$ and normalized by $\sigma$, plotted against $\tau$, 
for the nonlinear case ($r$), and for the Gaussian curvature approximation with the entropy correction ($r_{\rm EL}$) or without it ($r_{\rm L}$). Here, $N=6400$. 
For comparison, a plot of $(\sigma -\tau)/\sigma$ is shown.}
\label{fig:ren2}
\end{figure}

In the present work, all the nonlinearities of the problem have been taken into account: (i) the nonlinear expression of the membrane elementary area $dA$ (ii) the nonlinear expression of the membrane bending energy density $H^2 dA$, and (iii) the nonlinear measure correction $\mathcal{H}_\text{corr}$ dealing with the excess entropy of the Monge gauge. The Gaussian approximation, which is frequently used in basic calculations, consists in replacing the bending energy density $H^2 dA$ by $H_{\rm L}^2dA_{\rm p}$, where $H_{\rm L} =h_{xx}+ h_{yy}$ is the Laplacian approximation of the mean-curvature and $dA_p$ the projected elementary area, and in neglecting the entropy measure correction $\mathcal{H}_\text{corr}$. Indeed, at quadratic order, the measure correction becomes $\frac{1}{2}k_\mathrm{B}T\sum_{i,j}\theta_{ij}^2\simeq\frac{1}{2}k_\mathrm{B}T\sum_{i,j}(\nabla h)^2_{ij}$, which is of no consequences as it simply redefines the tunable parameter $\sigma$. In this subsection, we investigate how these two Gaussian approximations affect the surface tensions.
First, we make the Laplacian approximation, i.e.,
$H^2 dA$ is replaced by $H_{\rm L}^2dA_{\rm p}$ in eqn~(\ref{eq:sah}), and we keep the entropy correction term $\mathcal{H}_\text{corr}$. The corresponding results are denoted by the subscript EL in Fig.~\ref{fig:ren2}. Then we also remove the entropy correction term. The corresponding calculations are denoted by the subscript L.

We find that the internal tension $\sigma$ is not changed by these modifications. However, $r_{\rm EL}$ and $r_{\rm L}$ are significantly different from $r$, as shown in Fig.~\ref{fig:ren2}.
Using the Laplacian approximation while keeping the entropy correction term results in a fluctuation tension $r_{\rm EL}$ about three times larger than $\sigma$. Further removing the entropy correction yields a tension $r_{\rm L}$ very close to $\sigma$, in agreement with the well-known coincidence of $r$ and $\sigma$ at the Gaussian level.
Therefore, the effects of the measure correction and of the nonlinear curvature are opposite to each other: 
the former raises the value of $r$, the latter lowers it. 
Both of these effects need to be taken into account 
for a faithful description of the membrane fluctuations beyond the level of the Gaussian approximation, and the estimated $r$ deviates from the true value if either of them is missing.
Note that without the entropy correction term, $\mathcal{H}_{\rm corr}$, the nonlinear bending energy leads to an enhancement of the vertical motion of the sites, resulting in the collapse of the simulations.

\section{Conclusions\label{sec:concl}}

In this paper, we have examined the various surface tensions of membranes: frame tension $\tau$, internal tension $\sigma$, and renormalised ``fluctuation" tension $r$. We have compared them quantitatively using a Monte-Carlo simulation with control over both $\tau$ and $\sigma$, the latter being slaved to the former in order to keep the average membrane area $\langle A\rangle$ constant. We have also validated our results in the conjugated ensemble where the real membrane area $A$ is fixed (Appendix~\ref{ensembleAfixed}). We have investigated large systems in order to reach the thermodynamic limit.
Our model being a lattice model instead of a particle-based model, 
protrusion effects and other artificial elements are eliminated.  Gaussian approximation artefacts are also excluded because curvatures and areas are computed with exact formulas (although we employ the Monge gauge) and because we have corrected the excess entropy associated with the measure of the Monge gauge.
Moreover, because our simulated membrane does not exhibit rupture at extremely large $\tau$, all tensions from vanishingly small to very high can be investigated.

There have been two findings in our results. 
The first is that the residual internal tension $\sigma_0 = \sigma-\tau$ remains finite at large $N$ (large systems). 
While it is well known that $\sigma$ and $\tau$ are different, we have confirmed that the theoretical 
(albeit Gaussian) estimation of Ref.\citenum{Fournier2008}
predicts correctly and quantitatively the difference $\sigma-\tau$ from vanishingly small frame tensions to very large 
frame tensions. 
 
The renormalised ``fluctuation" tension $r$ has also been investigated, and
compared with the other two tensions. 
 We conclude
 $r\simeq\tau$ within an accuracy of $0.1\,k_{\rm B}T/a^2$ (estimated $\simeq\!10^{-6}\,{\rm J/m}^2$),
which is consistent with the proposition of Ref.\citenum{Cai1994a}. 

Note finally that the shape of the buckled membrane, described by elliptic functions, breaks the symmetry between the $x$ and $y$ directions and yields an anisotropy in the 
frame tension~\cite{Noguchi2011}.
How exactly the thermal fluctuations of the membrane modify
the mechanical buckling transition should be investigated in further studies.

\section*{Appendix}
\renewcommand{\thesubsection}{\Alph{subsection}}

\subsection{Simulation with nearly constant real area}
\label{ensembleAfixed}

In the body of this paper, the real area $A$ of the membrane is not fixed, but its average value $\langle A \rangle$ is controlled by the parameter $\sigma$ (much as the average number of particles in a grand-canonical ensemble is controlled by the chemical potential). In principle, we have checked the validity our results by studying the thermodynamic limit of increasing system sizes. It is nonetheless interesting to test our results by working in the ensemble in which the area $A$ is prescribed. This is the purpose of the present Appendix.

While it is very difficult to prescribe exactly the membrane area, because it is difficult 
to identify the sampling condition that satisfy 
this constraint, one can easily almost prescribe the area by using a quadratic potential of tunable strength $K_{\mathrm A}$. We thus worked with the Hamiltonian
\begin{equation}
\mathcal{H}^{*} = -\tau A_{\rm p} + \int_A \frac{\kappa}{2} H^2\ dA + \frac{K_{\mathrm A}}{2}(A-A_0)^2. 
\end{equation}

\begin{figure}
\centering
 \includegraphics[width=.9\linewidth]{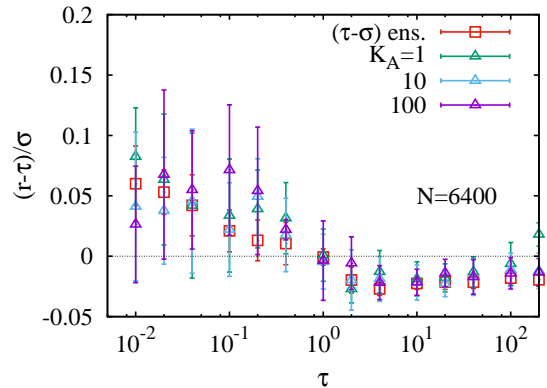}
\caption{
The deviation of renormalised tension $r$ from the 
frame tension $\tau$ is plotted in the form $(r-\tau )/\sigma$. 
The red points are pasted from Fig.~\ref{fig:ren}b for comparison. 
The results for three values of $K_{\mathrm A}$ (=1,\ 10,\ and 100) are shown by using symbols and error bars of different colours. For all the plotted data, the system size is $N=6400$.
Although we normalise the data by $\sigma$, 
no estimate is available in the present simulation; therefore, we use the values of Fig.~\ref{fig:ren}a as substitutes.
}
\label{fig:plot_dev2}
\end{figure}

We have performed the corresponding Monte-Carlo simulations exactly in the same manner as in the body of the paper. 
Figure \ref{fig:plot_dev2} shows the result of these additional simulations,
for $K_{\mathrm A} = 1$, 10 and 100, with $N=6400$, and compares them with the results of Fig.~\ref{fig:ren}. 
With increasing $K_{\mathrm A}$, the standard deviation of the real area decreases:
0.01\%,\ 0.001\%, and 0.0003\%, for $K_{\mathrm A} = 1,\ 10,$ and $100$, respectively. 
With larger values of $K_{\mathrm A}$, the fluctuations of the membrane area become suppressed so that unbiased equilibrium sampling is impossible.

\subsection{Entropy correction for membrane tilt}
\label{app:ang}

\begin{figure}
 \centering
 \includegraphics[width=.85\linewidth]{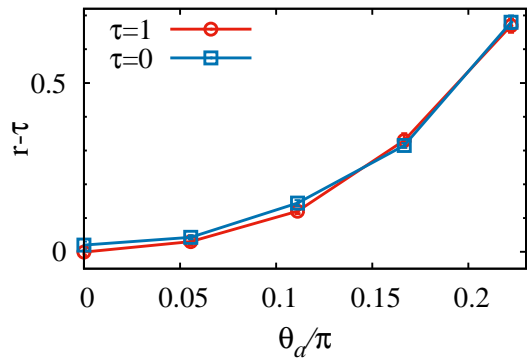}
  \caption{ 
Dependence of the fluctuation tension $r$ 
for a membrane whose projected plane
is entirely tilted by the angle 
$\theta_{\rm a}$, for $\tau=0$ and $\tau=1$, at $N=6400$.
}
\label{fig:ang}
\end{figure}

To check the reliability of the measure correction $\mathcal{H}_\text{corr}$, eqn~(\ref{eq:Hcor}),
we  performed simulations in which the entire membrane is tilted 
by an angle $\theta_{\rm a}$.
The periodic boundary condition along the $x$~direction is modified to $h(N_x,j)=h(0,j)+  N_x a \sin \theta_{\rm a}$.
The $x$~side length of the lattice is changed to $a \cos \theta_{\rm a}$
in order to maintain a square grid in the tilted projected plane.
The height spectrum is calculated for the tilted plane, of normal vector $(-\sin \theta_{\rm a}, 0, \cos \theta_{\rm a})$.
The fluctuation tension $r$ increases with increasing $\theta_{\rm a}$ (see Fig.~\ref{fig:ang}),
while $\sigma$ is independent of $\theta_{\rm a}$.
This implies that the excess entropy associated with the membrane tilt in  the Monge gauge is not completely removed.
However, the deviation is small for small $\theta_{\rm a}$,
and the overestimation of $r$ is less than 0.1 for $\theta_{\rm a}< 0.1\pi$.
In our simulation, at $\tau = 0$ we obtain
$\langle \cos\theta_{ij}\rangle =  0.98$,
i.e., the mean angle is $0.066\pi$. 
We conclude that the obtained values of $r$ are reliable within the accuracy of $0.1$ (in dimensionless units).

\subsection{Definition of the spectrum wavevectors}
\label{wavevectors}

In Sec.~\ref{sec:calc_r},
the renormalised tension $r$ is calculated by fitting 
the height spectrum $\langle |h (\bm{q})|^2\rangle$.
We use a sequence of instantaneous height configurations, 
and we perform averages after the simulation finishes. 
Because the projected area $A_{\rm p}$ changes during the simulations, 
the wavevectors on the reciprocal lattice are not fixed. They are given by
\begin{equation}
\bm{q} = (q_x,q_y) = \frac{2\pi}{\sqrt{A_{\rm p}} } (n,m),
\end{equation}
where $n$ and $m$ are integers. 
All these data are mapped to a one-dimensional wavevector $q=|\bm{q}|$. 
In plotting $\langle |h(q)|^2 \rangle$, the data of  $|h (\bm{q})|^2$ are averaged 
over close values of $q$ (the width of $q$ bins is $b \pi /\sqrt{N}$. 
Here, $b=1.2$ is taken) and over all the 
height configurations. 
 
We have confirmed the robustness of our estimation of $r$, 
by using a different set of wavevectors, defined by
\begin{equation}
\bm{q}' = \frac{2\pi}{\sqrt{\langle{A_{\rm p}} \rangle}} (n,m).
\end{equation}
 For each of integers 
$n,m$, the values of $\langle |h(\bm{q}')|^2\rangle$ 
is averaged over the sequence of instantaneous 
height configurations. The results obtained by these two methods do not essentially differ from 
one another: the differences in the value of $r$ are smaller than $0.025~k_\mathrm{B}T/a_0^2$
for all $\tau$, which is typically smaller than the error bars.

\section*{Acknowledgments}
The authors thank F. Schmid and H. Diamant for their comments.
This work was supported by the Core-to-Core Program ``Non-equilibrium 
dynamics of soft matter and information'' by the Japan Society for Promotion
of Science (JSPS) and also by a Grant-in-Aid for Scientific Research
on Innovative Areas ``Synergy of Fluctuation and Structure:
Foundation of Universal Laws in Nonequilibrium Systems'' 
(Grant No. 25103010). 
Numerical calculations were partly carried out on SGI Altix ICE 
8400EX System at ISSP, University of Tokyo.



\providecommand*{\mcitethebibliography}{\thebibliography}
\csname @ifundefined\endcsname{endmcitethebibliography}
{\let\endmcitethebibliography\endthebibliography}{}

\end{document}